\documentclass[11pt]{article}
\usepackage[top=1.1in, bottom=1.1in, left=1in, right=1in]{geometry}

\usepackage[ruled,vlined,linesnumbered]{algorithm2e}
\usepackage{url}
\usepackage{amsmath,amsfonts}
\usepackage{algorithmicx,algpseudocode}
\usepackage{caption}
\usepackage{subcaption}
\usepackage{graphicx}
\usepackage{textcomp}
\usepackage{xcolor}
\usepackage{multirow}
\usepackage{amsthm}
\usepackage{pgfplots}
\usepackage{listings}
\usepackage{lineno, blindtext}
\usepackage{pifont}
\usepackage{fancyvrb}
\usepackage{array}
\usepackage{siunitx}
\usepackage{changepage}

\theoremstyle{definition}
\newtheorem{definition}{Definition}

\newtheorem{claim}{Claim}
\def\BibTeX{{\rm B\kern-.05em{\sc i\kern-.025em b}\kern-.08em
    T\kern-.1667em\lower.7ex\hbox{E}\kern-.125emX}}

\newcommand{\comment}[1]{}

\AtBeginDocument{%
  \providecommand\BibTeX{{%
    \normalfont B\kern-0.5em{\scshape i\kern-0.25em b}\kern-0.8em\TeX}}}

\begin{document}

\title{A Symbolic Approach to Detecting Hardware Trojans Triggered by Don't Care Transitions
}

\author{Ruochen Dai \\ ruochendai@ufl.edu \\ ECE Department \\ University of Florida \and 
Tuba Yavuz \\ tuba@ece.ufl.edu  \\ ECE Department \\ University of Florida}

\date{}
\maketitle

\begin{abstract}
Due to the globalization of Integrated Circuit (IC) supply chain, hardware trojans and the attacks that can trigger them 
have become an important security issue. One type of hardware Trojans leverages the don't care transitions 
in Finite State Machines (FSMs) of hardware designs. 
In this paper, we present a symbolic approach to detecting don't care transitions and the hidden Trojans. 
Our detection approach works at both RTL and gate-level, does not require a golden design, and works in three stages. 
In the first stage, it explores the reachable states. In the second stage, it performs an approximate analysis 
to find the don't care transitions. 
In the third stage, it performs a state-space exploration from reachable states that have incoming 
don't care transitions to find behavioral discrepancies with respect to what has been observed in 
the first stage.
We also present a pruning technique based on the reachability of FSM states.
We present a methodology that leverages both  RTL and gate-level
for soundness and efficiency. 
Specifically, we show that don't care transitions must be detected at the gate-level, i.e., after synthesis has been performed, for soundness. However, under specific conditions, Trojan detection can be  performed more efficiently at RTL.
Evaluation of our approach on a set of benchmarks from OpenCores and TrustHub 
and using gate-level representation generated by two synthesis tools, Yosys and Synopsis Design Compiler (SDC), shows that our approach 
is both efficient (up to 10X speedup w.r.t. no pruning) and precise (0\% false positives) in detecting don't care transitions and the Trojans that leverage them. 
Additionally, the total analysis time can achieve up to 3.40X (using Yosys) and 2.52X (SDC) speedup when 
synthesis preserves the FSM structure and the Trojan detection is performed at RTL.
\end{abstract}




\section{Introduction}

Globalization of the Integrated Circuit (IC) supply chain and the role of Third-party Intellectual Property (3PIP) 
in SoC development opened up possibilities for various attacks in the hardware domain. One type of attack is injecting malicious logic, which is called a hardware Trojan, into hardware designs or implementations.  The Trojan is generally designed to be triggered on a certain input sequence or event to perform its malicious activity, which can be harming the circuit, changing the intended behavior, or leaking secret information.

Although the distributed nature of hardware development workflow is an important factor for this type of attack surface, 
the way a hardware component is designed and implemented can also have a role in the success of such attacks.
An important type of vulnerability in hardware design is the don't care transitions in the Finite State Machines (FSMs). 
An FSM implements sequential control logic of a hardware component. When the transition function of an FSM is not defined completely \cite{DQ14}, there may exist some don't care transitions from some states. 
Although don't care transitions may provide opportunities for optimizing the synthesis,
an attacker can exploit such transitions as a trigger mechanism for the malicious payload  of a Trojan  \cite{DQ14}.

This paper presents a self-referencing (golden design free) Trojan detection at Register Transfer Level (RTL) as well as at gate-level and targets 
Trojans that are triggered by don't care transitions. 
These features make our approach applicable to four out of the seven types of attack models described in \cite{XFJ16}, 
in which the Third-party IP (3PIP) vendors, who typically deliver the IPs at RTL, are considered untrusted. 
These attack models consist of untrusted 3PIP vendor, commercial off-the-shelf component, 
untrusted design house, and fabless SoC design house.

\begin{figure*}[th!]
\begin{center}
\includegraphics[width=0.8\textwidth]{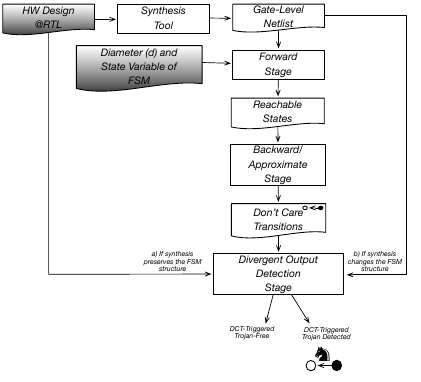}
\caption{Overall workflow of the symbolic Trojan detection approach. Gray-filled boxes represent the inputs. Black-filled and white-filled circles represent the unreachable and reachable states of the FSM, respectively.}
\label{fig:workflow}
\end{center}
\end{figure*}

The overall workflow of our approach is shown in the Figure \ref{fig:workflow} and leverages symbolic execution \cite{King76}, a software testing technique that can 
potentially achieve high coverage of the system under analysis. 
We have implemented our approach on top of the KLEE symbolic execution engine \cite{CDE08}.
We follow the approach in \cite{ZDH18} and translate Verilog designs and their synthesized gate-level netlists into their C++ implementations using Verilator, which 
is a simulator for hardware designs \cite{Verilator}. So, we use the KLEE symbolic execution engine with don't care 
transition and Trojan detection extensions on the C++ implementations generated by Verilator.

The presented approach works in three stages. In the first stage, we use symbolic execution to 
perform forward reachability analysis on the synthesized gate-level netlist up to a given number of clock cycles, which should be at least the diameter of the FSM under analysis. 
This stage is performed to compute the set of reachable states and input/output behavior while exercising all possible input sequences.
In the second stage, we use an approximate analysis to find out the predecessors of reachable states. 
The goal of the second stage is to identify the don't care transitions by exercising all possible source states 
and inputs. This stage uses the reachable states computed in stage one to identify unreachable source states and 
reachable destination states. 

The reason for performing the first and the second stages on the gate-level
 netlist instead of the RTL design is that the synthesis tools may introduce additional don't care 
 transitions or change the FSM structure. 
 Considering the fact that the Trojan attacks targeted in this paper would  
 exploit the don't care transitions that exist in the synthesized circuit, this is important to achieve soundness for detecting Trojans that are 
 triggered by don't care transitions.  
 
Finally, in the third stage we perform another forward analysis up to a given number of clock cycles to 
detect any input-output behavior deviations for states that are reachable from states with incoming don't care transitions.
So, our approach reports both the don't care transitions and the Trojans hidden at such transitions, if any, to help the designers harden their designs and evaluate the trustworthiness of 3PIPs. Our approach achieves optimized performance 
in two ways. The first one leverages a pruning approach that reduces the analyzed state space 
while guaranteeing sufficient coverage in inferring 
the set of reachable states. 
The second one leverages the structural properties of the synthesed FSM. If the synthesis tool preserves the FSM in terms of the states and the transitions between the states then the third stage can be performed at RTL. Our experiments show that 
this provides a more efficient detection while guaranteeing soundness. However, 
when the synthesis tool does not preserve the structure of the original FSM then the third stage must be performed at the synthesized gate-level 
netlist to achieve soundness with a performance cost. 

Our tool will be publicly released along with 
the benchmarks from OpenCores \cite{OpenCore} and those that were generated by adapting benchmarks from Trust-hub \cite{TrustHub} and \cite{IMA_ADPCM}.

This paper is organized as follows. Section \ref{sec:relwork} places our work within the context of related work on finite state machine analysis and Trojan detection. 
Section \ref{sec:approach} presents the technical details of our approach. 
Section \ref{sec:eval} presents evaluation of our approach on various benchmarks.
Section \ref{sec:discussion} discusses the limitations of our approach.
Section \ref{sec:conc} concludes with directions for future work.

\section{Related Work}
\label{sec:relwork}
In this section, we discuss two groups of related work. The first group includes work that detect don't care transitions. 
The second group includes work that detect hardware Trojans.

\paragraph{Finite State Machine (FSM) Analysis}

The Netlist Analysis Toolset (NETA)  \cite{MPZ19} has been designed to help IP users extract FSMs from  the
low gate-level logic. However, this work does not identify any don't care transitions.

In \cite{DQ14}, the authors consider don't care vulnerabilities in FSMs by using state reachability as a trust metric and by defining \emph{incompletely specified FSMs} as those with unspecified next-states or output functions. 
Untrusted implementation of an FSM is detected by checking the existence of discrepancies between the 
original and the implemented FSMs in terms of the reachable states and the predecessors. 
The authors also discuss possible attacks that leverage the don't care transitions in unsafe implementations.

The AVFSM framework  \cite{NXY16} extracts a state transition graph (STG) from a gate level netlist using an 
Automatic Test Pattern Generation (ATPG) approach. Don't care states and transitions are detected manually by 
comparing the extracted STG and the one that is generated from the RTL design by an EDA tool. 

In \cite{FM17}, equivalence checking between an FSM specification and 
its gate level implementation is used to detect anomalies in the implementation. 
Both the specification and the implementation are represented as polynomials and 
symbolic algebra is used for equivalence checking.

\paragraph{Hardware Trojan Detection}

Various heuristics on how often Trojans get activated have been used to detect Trojans.
These heuristic based approaches include identification of unused circuits \cite{HFK10}, 
suspicious signals  \cite{ZT11}, suspicious wires \cite{WSS13}, and 
dedicated triggers \cite{ZYW15}. 
 
Formal verification approaches to Trojan detection require either functional specifications \cite{RSK14,GDJ15} 
or security relevant specifications \cite{LJM12,RVK15}. 
    
Self-referencing techniques  \cite{NWD11,ZXT13,LHM14,ZZ17,HNW17,XR18,XBL19}, which eliminate the 
need for a golden chip, leverage various physical characteristics of the circuits to detect Trojans during post-silicon analysis.
        
Dangerous don't care signals are detected in \cite{FKC15} by checking existence of distinguishing input sets that 
yield different observable output values. The analysis performs combinational equivalence checking on the attacker observable outputs between two copies of the same RTL design that differ only in the assignments of the don't care signals.

To our knowledge, our don't care transition identification and Trojan detection approach is the first self-referencing technique at RTL as well as at gate-level, thanks to leveraging an approximate symbolic exploration of the state space. Although our approach focuses on the don't care transition triggered Trojans, unlike the heuristic based approaches \cite{HFK10,ZT11,WSS13,ZYW15}, it does not rely on implementation details. 
Also, unlike the formal verification based approaches, our approach does not require any specifications.
Finally, unlike \cite{FKC15}, our approach considers sequential semantics as it computes the set of reachable states.

\section{Approach}
\label{sec:approach}

In this section, we present the technical details of our approach for detecting Trojans hidden at don't care transitions.
Section \ref{sec:prel} presents some fundamental notations about Finite State Machines (FSMs) with inputs and outputs along with some observations that we leverage in our detection approaches.
Section \ref{sec:dont} presents our two-staged approach for don't care transition detection and discusses various symbolic exploration approaches. 
Section \ref{sec:trojan} presents our three-staged approach for detecting Trojans hidden at don't care transitions 
and explains how symbolic execution is utilized in all the three stages.

\subsection{Preliminaries}
\label{sec:prel}

\begin{definition}[Mealy Machine]
A Mealy machine $M$ is a 6-tuple $(S, S_0, \Sigma, \Lambda, T, G)$, where $S$ is the set of states, $S_0$ is the 
set of initial states, $\Sigma$ and $\Lambda$ are the sets of input and output alphabets, respectively, $T: S \times \Sigma \rightarrow S$ is a transition function, and $G: S \times \Sigma \rightarrow \Lambda$ is an output function.
\end{definition}

\begin{definition}[Reachability]
A state $s \in S$  is reachable in $M$ if either $s$ is in $S_0$ or there exists a sequence of inputs 
$i_1, ..., i_k$ such that starting from some initial state $s_0 \in S_0$, a sequence of transitions visit $s_0,s_1,...,s_k$, where $s_k=s$, according to the 
transition function $T$, i.e., $\exists k > 0. \ \exists s_0 \in S_0. \ \exists i_1, i_2, ..., i_k.. \ s_k=s \ \wedge \ \forall 0 \leq j < k. \ T(s_j,i_{j+1})=s_{j+1}$.
A state $s$ is unreachable if it is not reachable.
\end{definition}

\begin{definition}[Bounded Reachability]
Let $R(M)$ denote the reachable states in $M$. For a given integer $j \geq 0 $, $R(M)_j$ denotes the set of reachable states 
that are reachable in at most $j$ steps. $R(M)_0=S_0$, and there exists $j \geq 1$ and $s_0 \in S_0$, $s \in R(M)_j \leftrightarrow 
(s \in R(M)_{j-1} \ \vee \ \exists i_1, i_2, ...., i_j. \ \forall 0 \leq k < j. \ T(s_k,i_{k+1})=s_{k+1}  \ \wedge \ s_j=s)$.
\end{definition}

\begin{definition}[Diameter and Rank]
The maximum distance of a state $s$ in $M$ is the longest Hamiltonian path 
from some initial state in $S_0$ to $s$. The diameter of $M$, $D(M)$, is the largest maximum distance among all the reachable states in $M$. 
For a given reachable state $s$ in $M$, the rank of $s$ is defined as the smallest $j$ such that $s \in R(M_j)$.
\end{definition}

\begin{definition}[Cut]
Let $M^{S'}$ denote a cut of $M=(S, S_0, \Sigma, \Lambda, T, G)$ with respect to a set of initial states $S'$, i.e.,  
$M^{S'}=(S, S', \Sigma, \Lambda, T, G)$, where $S' \subseteq S$. 
\end{definition}

\begin{claim}[Monotonicity of Bounded Cut Reachability]
\label{claim:mbcr}
Given a Mealy Machine $M$ and a state $s$ that is reachable in $M$, the set of reachable states from $s$ 
does not shrink with increasing values of bounds, i.e., $\forall j.k. \ 0 \leq k < j. \ R(M^{\{s\}})_k \subseteq R(M^{\{s\}})_j$.
\begin{proof}
Follows from the monotonicity of the successor function.
\end{proof}
\end{claim}

\SetKwInOut{Parameter}{Parameters}
\begin{algorithm}[th!]
\SetAlgoLined
\KwIn{$M=(S, S_0, \Sigma, \Lambda, T, G)$: Mealy Machine, $d$: Diameter}
\KwOut{Set of don't care transitions} 
$s_0 \gets init(P)$\;
$RS \gets \{s_0\}$\;
 \For{$i$: 1 to $d$}{
  $RS \gets RS \ \cup \ \{s' \ | \ s \in RS \ \wedge \ \exists I. T(s,I)=s'\}$
 }
 $Trans \gets \{ (s_1,s_2) \ | \ s_2 \in RS \wedge \exists I. T(s_1,I)=s_2\}$\;
 $DCT \gets \{ (s_1,s_2) \ | \ s_1 \not \in RS \ \wedge \ (s_1,s_2) \in Trans\}$\;
 return $DCT$\;
 \caption{Computing the set of don't-care transitions in a Mealy Machine $M$.}
 \label{alg:twostage}
\end{algorithm}

\begin{definition}[Input Partition]
\label{def:part}
Let $M_I=(S, S_0, \Sigma, \Lambda, T', G')$ denote the partitioning of $M$ with respect to a sequence of 
input sets $I=I_1,I_2,..,I_k$, where $k$ is at most $D(M)$, $\forall 1 \leq j \leq k. \ I_j \subseteq \Sigma$, and 
\begin{equation}
\begin{split}
s_0 \in S_0  \ \wedge \ 
\forall 1 \leq j \leq k. \\ 
((i_j \in I_j \ \wedge 
T(s_{j-1},i_j)=s_j \ \wedge \ G(s_{j-1},i_j)=o_j) \ \leftrightarrow  
(T'(s_{j-1},i_j) =s_j \ \wedge  \ G'(s_{j-1},i_j)=o_j)) \ \wedge \\
(i_j \not \in I_j \ \leftrightarrow (T'(s_{j-1},i_j) = undef \ \wedge \ G'(s_{j-1},i_j)=undef))
\end{split}
\end{equation}
\end{definition}

\begin{claim} 
\label{claim:subset}
The set of reachable states in a partition of $M$ with respect to $I$ is a subset of or equal to 
the set of reachable states in $M$, i.e., $R(M_I) \subseteq R(M)$.
\begin{proof} Follows from the fact that the initial states of $M_I$ is the same as the initial states of $M$ and that the transitions in $M_I$ is a subset of the transitions in $M$.
\end{proof}
\end{claim}


\begin{definition}[Predecessors]
The set of predecessors of state $s$ for input $i$, $Pred(s,i)$, in $M$ is the set of predecessor states that can reach $s$ in one step on input $i$, i.e., $Pred(s,i) = \{ s' \ | \ \exists s' \in S. \ T(s',i)=s\}$. 
\end{definition}

\begin{definition}[Don't Care Transition]
A don't care transition from $s'$ to $s$ 
on input $i$ in $M$ exists if $s'$ is unreachable in $M$, $s$ is reachable in $M$, and there is a transition from 
$s'$ to $s$, i.e., $s' \not \in R(M) \ \wedge s \in R(M) \ \wedge \ T(s',i)=s$.
\end{definition}

\begin{figure}[th!]
\begin{footnotesize}
\begin{Verbatim}[numbers=left,xleftmargin=5mm]
always @ (posedge clock or posedge reset)
begin 
  if (reset) ..
  else
  begin 
    case (pcmSq) // waiting for a new input sample 
      `IDLE:
         if (inValid) 
            begin 
                ...
                pcmSq <= `SIGN;
            end 
         else ...
      `SIGN:// check the difference sign and 
               // set PCM sign bit accordingly 
         begin 
             ...
             pcmSq <= `BIT2;
         end 
      `BIT2:
          begin 
              ...
              pcmSq <= `BIT1;
	  end 
      `BIT1:
          begin 
              ...
              pcmSq <= `BIT0;
          end 
      `BIT0:
          begin 
              ...
              pcmSq <= `DONE;
          end 
      `DONE:
          begin 
              ...
              pcmSq <= `IDLE;
          end 
       // unused states 
       default:	pcmSq <= `IDLE;
     endcase 
   end 	
end	
\end{Verbatim}
\end{footnotesize}
\caption{A snippet of the Verilog implementation of the FSM in the IMA ADPCM benchmark.}
\label{fig:imafsm}
\end{figure}

\begin{figure*}[th!]
\begin{center}
\includegraphics[width=0.65\textwidth]{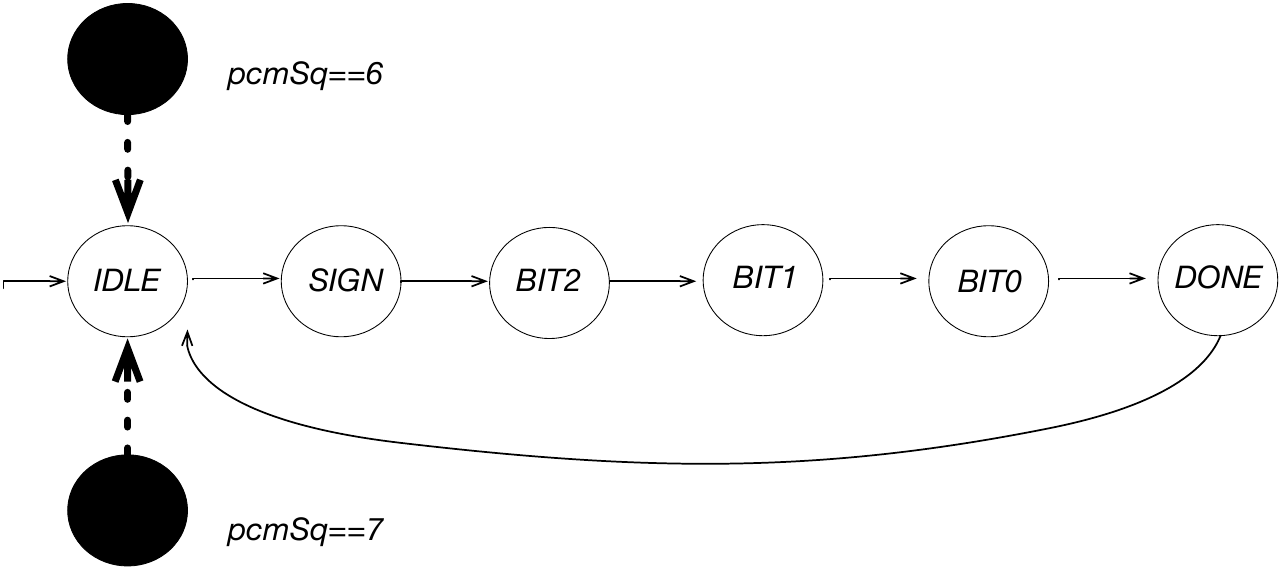}
\caption{The FSM that corresponds to the RTL design in Figure \ref{fig:imafsm}. White-filled  and black-filled circles represent the reachable and unreachable states, respectively. Dashed lines represent don't care transitions.}
\label{fig:fsmex}
\end{center}
\end{figure*}

Figure \ref{fig:imafsm} shows a snippet of Verilog code for the FSM part of one of our benchmarks, IMA ADPCM, 
which implements an audio compression algorithm \cite{IMA_ADPCM}. 
The {\tt pcmSq} register stores the FSM state and it is three bits wide. However, only values 0-5 are reachable states 
as the lines 11, 18, 23, 28, 33, and 38 indicate. The {\tt default} case at line 41 allows transitioning from unreachable 
states, with values 6 and 7, to one of the reachable states, {\tt `PCM\_IDLE}, which is defined as 0. 
The FSM with reachable and unreachable states and the don't care transitions of this design are depicted in Figure \ref{fig:fsmex}.
Such unreachable states can be realized through a physical attack and by exploiting don't care transitions that take the 
FSM from unreachable states to reachable ones, the attacker can violate safety and security properties.

\subsection{Detecting Don't Care Transitions}
\label{sec:dont}

In this section, we first introduce our approach at a high-level in Section \ref{sec:high}.
We briefly introduce a background on Symbolic Execution in Section \ref{sec:background}.
We discuss detecting don't care transitions using symbolic execution in Section \ref{sec:twostage}. 

\subsubsection{High-level Approach}
\label{sec:high}

\begin{figure}[th!]
\centering
\begin{subfigure}{.2\textwidth}
  \centering
  \includegraphics[width=\linewidth]{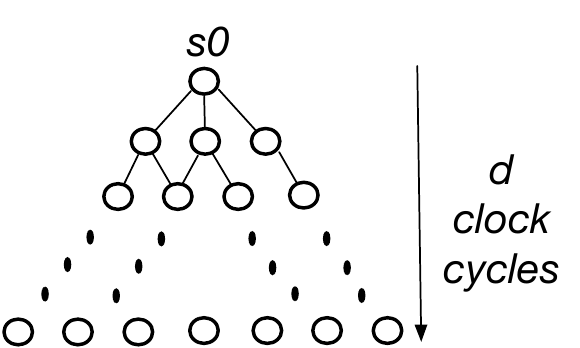}
  \caption{Forward Stage}
  \label{fig:sub1}
\end{subfigure}%
\begin{subfigure}{.2\textwidth}
  \centering
  \includegraphics[width=\linewidth]{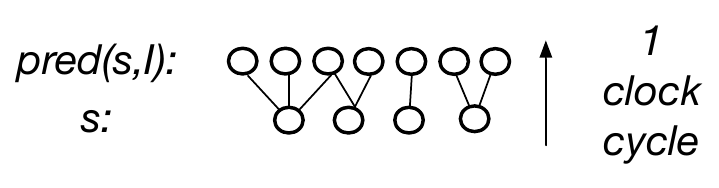}
  \caption{Backward Stage}
  \label{fig:sub2}
\end{subfigure}
\caption{Two stages of don't care transition detection.}
\label{fig:twostage}
\end{figure}

We assume that the diameter of the 
Mealy Machine $M$ under analysis is known and denoted by $d$.
Figure \ref{fig:twostage} shows our two stage approach.
 The first stage consists of a forward analysis, which computes the set of reachable states in $d$ steps.
The second stage uses a backward analysis, which compute the predecessors of every state in $M$. 
Algorithm \ref{alg:twostage} shows the high-level algorithm of our approach.

\subsubsection{Background on Dynamic Symbolic Execution}
\label{sec:background}

Dynamic symbolic execution is a static program analysis technique that can reason about symbolic 
inputs. The word ``dynamic'' refers to the fact that concrete and symbolic values can be mixed 
on an execution path \cite{CS13}. 
Dynamic symbolic execution has two major flavors: concolic and execution-tree generation based. 
A symbolic execution engine typically interprets the instructions of an 
intermediate language, such as the LLVM IR \cite{LLVM,LA04}, so that expressions that involve symbolic 
values are manipulated according to the semantics of the instruction. 
In this paper, we focus on the execution-tree generation based approach.
When interpreting conditional branch instructions with symbolic branch conditions, a symbolic 
execution engine checks the satisfiability of the branch condition for each target using an SMT solver 
and to simulate each feasible target it generates a separate path. 
On each path it conjoins the symbolic branch conditions to generate the {\em path constraint}.
So, the symbolic execution 
execution engine generates a tree of symbolic execution paths or states, where the internal nodes with multiple children 
denote branching points and each leaf node denotes a completed execution corresponding to an 
equivalence class of the input space. 
A challenge in symbolic execution is the well-known path explosion problem as the tree of executions 
may grow exponentially with the increasing number of branching instructions. 
So, symbolic execution is typically configured to run up to some timeout value.

\subsubsection{Finding Don't Cares Using Symbolic Execution}
\label{sec:twostage}

In this section, we explain how we use symbolic execution to perform the two-stage analysis 
that we present in Figure \ref{fig:twostage} and Algorithm \ref{alg:twostage}.
We have implemented our approach on top of the KLEE symbolic execution engine.
We have used Verilator to translate the RTL design as well as the synthesized gate-level netlist into C++, and we used the clang compiler 
to generate the LLVM bitcode from the C++ code. 
One of the strengths of symbolic execution is the ability to explore the input space in a systematic way.

In the forward stage, we label the input variables, e.g., {\tt inValid} in Figure \ref{fig:imafsm}, as symbolic in each of the $d$ number of steps. 
This makes sure that each step gets a fresh symbolic variable and that the input values for each step are 
independent. Therefore, all sequences of input equivalence classes are explored within a time bound. 
We record the FSM state reached in each symbolic execution state in a global metadata structure that 
represents the set of reachable states.

\begin{figure*}[th!]
\centering
\begin{subfigure}{.4\textwidth}
  \centering
  \includegraphics[width=\linewidth]{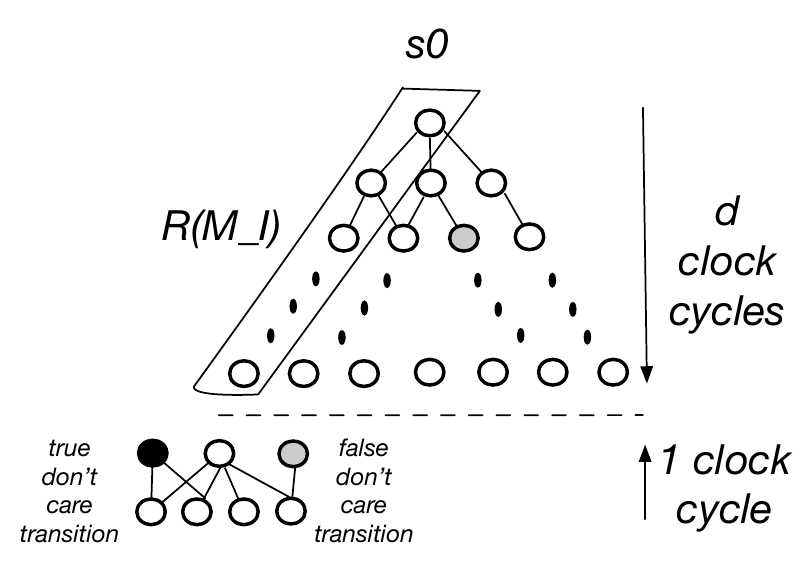}
  \caption{The partial exploration approach.}
  \label{fig:sub1}
\end{subfigure}%
\begin{subfigure}{.37\textwidth}
  \centering
  \includegraphics[width=\linewidth]{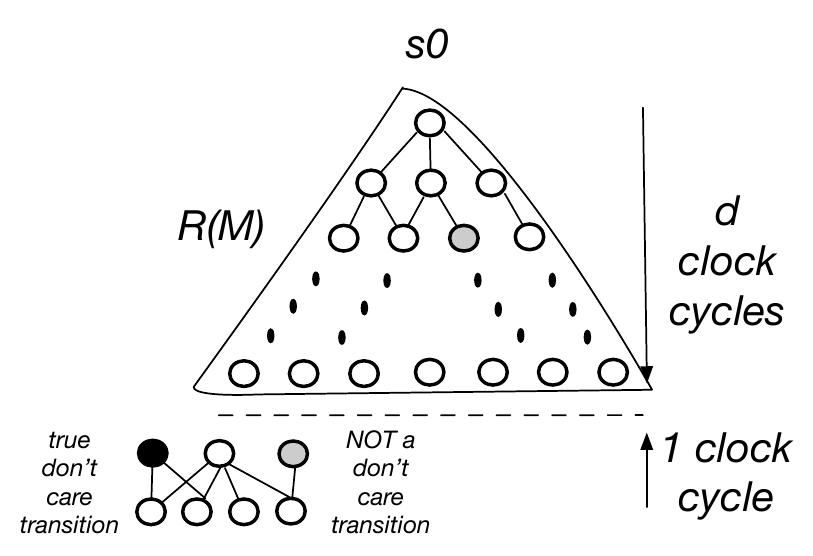}
  \caption{The full exploration approach.}
  \label{fig:sub2}
\end{subfigure}
\caption{Two approaches for detecting don't care transitions using symbolic execution. $d$ is the diameter of the 
analyzed Mealy machine $M$. $I$ denotes an input sequence. $R(M)$ and $R(M_I)$ denote the set of reachable 
states in $M$ and in a partition of $M$ w.r.t. the input sequence $I$, respectively. Black nodes represent unreachable states.}
\label{fig:impl}
\vspace{-0.5cm}
\end{figure*}

In the second stage, we perform symbolic execution for one clock cycle.
{\em Also, we label both the input variables and the FSM's state variable as symbolic}.
Although symbolic execution is a type of forward analysis, by marking the state variable, e.g., {\tt pcmSq} in 
 Figure \ref{fig:imafsm}, as symbolic 
 we are able to simulate the backward analysis, which finds the predecessors of every state. 
Therefore, this stage is an approximate execution as opposed to the precise execution in the forward stage.
This means that it is possible that the source state is an unreachable state, e.g., {\tt pcmSq == 6}. 
Although an approximate analysis like this may lead to false positives when reasoning about the 
behavior that the circuit may demonstrate, at the same time, it reveals the don't care transitions, the main goal of this stage, as 
a don't care transition is from an unreachable state to a reachable state. 
So, in this approximate stage, we use the set of reachable states computed in the forward stage to
check whether the source state is an unreachable state and the destination state is a reachable state.
Such transitions are identified as don't care transitions and are reported to the user.

To implement the two stage approach for don't care transition detection using symbolic execution, we have mainly 
two options as shown in Figure \ref{fig:impl}.
Figures \ref{fig:impl}.a and \ref{fig:impl}.b both show the symbolic execution trees for a Mealy machine under analysis.
In these figures, the circles with black color represents the unreachable states and all other states are the reachable ones.

The first option of using symbolic execution is depicted in Figure \ref{fig:impl}.a, which considers each symbolic execution path as a partition of the Mealy machine $M$ with respect to an input sequence $I$ (see Definition \ref{def:part}) as long 
as the diameter of $M$ and computes the set of reachable states that have been 
encountered on that path: $R(M_I)$. However, as indicated by Claim \ref{claim:subset}, the set of 
reachable sets observed in a partition may miss some of the reachable states as illustrated by the 
gray colored state in Figure \ref{fig:impl}.a. As a result, it is possible to identify some states as unreachable 
while they are reachable on a different symbolic execution 
path that exercises a different input sequence. 
Therefore, with the partial exploration of the reachable states, it is possible to have false positive don't care transitions. 
For the FSM in Figure \ref{fig:imafsm}, if at each clock cycle on a symbolic execution path the input {\tt inValid} evaluates to false, only {\tt `PCM\_IDLE} 
would be considered 
as a reachable state as the FSM state will not change (line 13) and the other 4 reachable states would be incorrectly considered 
as unreachable.

The second option of using symbolic execution is depicted in Figure \ref{fig:impl}.b, which 
computes the set of reachable states by the full exploration of the input space for the Mealy machine $M$ 
and up to the diameter of $M$. Unlike the partial approach, this technique can in theory guarantee precision of the 
set of reachable states. This means that all the reported don't care transitions would be true positives.
However, due to the well-known path explosion problem of symbolic execution, it may 
not be feasible to explore the full state space. 

Symbolic execution engines, such as KLEE, typically implement various path exploration strategies to deal with the path explosion problem. In addition to the classical search algorithms such as Depth-First Search (DFS) and Breadth-First Search (BFS), random and coverage based path exploration strategies exist. 

In this paper, we propose and analyze three variations of don't care transition detection based on the two main approaches and the path exploration algorithms. The three approaches are as follows:

\paragraph{The partial exploration approach} This is the approach that is depicted in Figure \ref{fig:impl}.a and uses 
a single symbolic execution path to compute the set of reachable states. It also uses the default path exploration 
strategy of KLEE, which is a combination of random and coverage based.

\paragraph{BFS-based exploration} This is the approach that is depicted in Figure \ref{fig:impl}.b and performs 
a full exploration of the symbolic execution tree in the forward stage using the BFS exploration 
strategy.

\paragraph{BFS-based exploration with pruning} This is a variant of the full exploration approach that is 
depicted in Figure \ref{fig:impl}.b. However, we configure the symbolic execution engine to use the BFS exploration 
strategy with a pruning technique that leverages Claim \ref{claim:mbcr}, which implies that if a state has already been 
explored in BFS then exploring the state at the same or lower layer in the symbolic execution tree (same or higher rank) would not yield finding more reachable states from that state. Therefore, in this approach we kill symbolic execution states 
with a state of the Mealy machine that has been encountered before. 

\subsection{Detecting Trojans Triggered by Don't Cares}
\label{sec:trojan}

In this section, we present our overall approach for don't care transition triggered Trojan detection. 

\begin{figure}
\begin{footnotesize}
\begin{Verbatim}[numbers=left,xleftmargin=5mm]
always @(posedge clock or negedge reset)
  begin
    if (reset) trojan_state <= trojan_idle;
    else begin          
      case (trojan_state)                
         trojan_idle: begin 
           if (pcmSq == 3'd6 || pcmSq == 3'd7)
              trojan_state <= trojan_active;
           else 
              trojan_state <= trojan_idle;
           end  
         trojan_active: begin 								
           if (pcmSq == 3'd0)
              trojan_state <= trojan_work;
           else 
              trojan_state <= trojan_active;
	   end 
         trojan_work: trojan_state <= trojan_work;
      endcase
    end
end
always @(posedge clock or negedge reset)
    if (reset) trojan_ena <= 1'b0;
    else if (trojan_state==trojan_work) 
       trojan_ena <= 1'b1;
\end{Verbatim}
\end{footnotesize}
\caption{A Trojan that gets triggered upon a don't care transition within the FSM of IMA ADPCM benchmark.}
\label{fig:trojantrigger}
\end{figure}

Figure \ref{fig:trojantrigger} shows the Trojan trigger logic that we implemented as an FSM and added to the IMA ADPCM benchmark. The Trojan is initialized to be in the {\tt trojan\_idle} state (line 3). 
When the FSM in Figure \ref{fig:imafsm} reaches one of the states that are normally unreachable (states with values 6 or 7 as shown at line 7), possibly through a physical attack such as voltage glitch, the Trojan FSM transitions to the {\tt trojan\_active} state (line 8).
However, to trigger the Trojan, the FSM waits to detect transitioning to a reachable state from an unreachable state, i.e., 
detecting a don't care transition, which is detected at line 13 and at line 14 the Trojan FSM's state is updated to 
{\tt trojan\_work}. Finally, the always block at lines 22-25 sets the {\tt trojan\_ena} line to high to denote that the Trojan payload can now be activated.

\begin{figure}
\begin{footnotesize}
\begin{Verbatim}[numbers=left,xleftmargin=5mm]
always @ (posedge clock or posedge reset)
begin 
   if (reset)
      begin 
        outPCM <= 4'b0;
        outValid <= 1'b0;
      end 
   else if (pcmSq == `PCM_DONE)
      begin 
        outPCM <= prePCM;
        outValid <= 1'b1;
   end 
   else if (trojan_ena)
      //trojan payload, stuck at 1 fault
      outValid <= 1'b1;
   else 
      outValid <= 1'b0;
end 
\end{Verbatim}
\end{footnotesize}
\caption{The Trojan payload for the IMA ADPCM benchmark.}
\label{fig:trojanpayload}
\end{figure}

Figure \ref{fig:trojanpayload} shows the Trojan payload (lines 13-15), which sets the output line {\tt outValid} to high when it is supposed to be low according to the original design.

\SetKwInOut{Parameter}{Parameters}
\begin{algorithm}[th!]
\SetAlgoLined
\KwIn{$P$: HW Design, $SVar$: FSM State Variable, $Input$: Input Variables, $Output$: Output Variables, $d$: Diameter}
\KwOut{Divergent behavior of the Trojan in $P$, if any}
$(RS, RBS) \gets$ $SymEx(P,\{Init(P)\},Input,Output,SVar,d,Reach)$\;
$(Trans, SymStates) \gets SymEx(P,True,Input \cup \{SVar\},Output,SVar,1,States)$\;
$DCT \gets \{ (s_1,s_2) \ | \ s_1 \not \in RS \ \wedge \ (s_1,s_2) \in Trans\}$\;
 \If{$DCT = \emptyset$}{
    {\bf print} "No don't care transitions and relevant Trojans"\;
    {\bf return} $\emptyset$\;
}
$DBS \gets \emptyset$\;
$Dest \gets \{ symst \ | \ Project(symst,SVar) = s_2 \ \wedge \ \exists s_1. \ (s_1,s_2) \in DCT\}$\;
\For{each $s \in Dest$}{
    $(RS', RBS') \gets SymEx(P,\{s\},Input,Output,SVar,d,Reach)$\;
    $DBS \gets DBS \ \cup \ RBS' \ \setminus \ RBS$\; 
}
\If{$DBS \not = \emptyset$}{
  {\bf print} "Trojan detected"\;
}
{\bf return} $DBS$\;
\caption{Detecting Trojans hidden at don't care transitions of a HW Design that is represented by program $P$.}
 \label{alg:threestage}
\end{algorithm}

Algorithm \ref{alg:threestage} shows all stages of our Trojan detection approach. 
In the first phase (line 1), the algorithm performs the full exploration of the symbolic execution tree for 
$d$ clock cycles as shown 
in Figure \ref{fig:impl})b  to compute 
the set of reachable states, $RS$, and the intended set of behaviors, $RBS$, in terms of a tuple consisting 
of the source state, $s$, the destination state, $s'$, enabling input, $I$, and the generated output $O$.

\SetKwInOut{Parameter}{Parameters}
\begin{algorithm}[th!]
\SetAlgoLined
\KwIn{$P$: HW Design, $Init$: Initial State(s), $Input$: Input Variables, $Output$: Output Variables, $SVar$: FSM State Variable, $N$: Number of Clock Cycles, $Type$: Metadata Type}
\KwOut{Metadata}
$M$: Metadata\;
$M \gets (\{Project(Init,SVar)\},\emptyset)$\;
$Init.numSteps \gets 0$\;
$Active \gets Init$\;
\While{$Active \not = \emptyset$ and $\exists s \in Active$ s.t. $s.numSteps < N$}{
   $symState \gets chooseNext(Active)$\;
   \If{$symState.numSteps < N$}{ 
       Symbolize $Input$ in $symState$\;
       $succs \gets  SymExForOneClockCycle(symState)$\;
       $Active  \gets Active \ \setminus \ \{symState\}$\;
       \For{each $s \in succs$ s.t. $s$ not terminated}{
           $s_1 \gets Project(symState,SVar)$\;
           $s_2 \gets Project(s,SVar)$\;
           \If{$Type$ is Reach}{
              $I \gets Project(s,Input)$\;
              $O \gets Project(s,Output)$\;
              $M.RS \gets M.RS \ \cup \ \{s_2\}$\;           
              $M.RBS \gets M.RBS \ \cup \ \{ (s_1,s_2,I,O)\}$\;
           }
           \If{$Type$ is States}{
              $M.Trans \gets M.Trans \ \cup \ \{(s_1,s_2)\}$\;
           }
           $Active \gets Active \ \cup \ \{s\}$\;
           $s.numSteps \gets symState.numSteps + 1$\; 
       }
   }
}
\If{$Type$ is Reach}{
   {\bf return} $(M.RS,M.RBS)$\;
}
\If{$Type$ is States}{
    $M.SymStates \gets Active$\;
    {\bf return} $(M.Trans,M.SymStates)$\;
}
 \caption{Symbolic Execution Algorithm ({\bf SymEx}) as adapted for a HW design.}
 \label{alg:symex}
\end{algorithm}

In the second phase (lines 2-7), the algorithm computes the set of don't care transitions, $DCT$, based on the 
set of reachable states and the transition function. Any transition that starts from an unreachable state and 
ends at a reachable state is included in $DCT$.
If there are no don't care transitions, then the algorithm terminates as there cannot be any Trojans that would be triggered 
with don't care transitions. 
Otherwise, the algorithm moves to the third phase (lines 8-13), to identify any divergent behavior that the design can demonstrate once triggered by the don't care transition.  

Algorithm \ref{alg:symex} presents an abstracted version of the symbolic execution algorithm as adapted to a hardware design, at RTL or gate-level. 
We assume that $P$ represents a program that simulates the  behavior of a hardware design. In our approach,
$P$ is generated by Verilator from a given Verilog design.
The algorithm gets the set of initial states, the input variables, the output variables, the variable in $P$ that 
represents the FSM state, the number of clock cycles, and the type of metadata to be computed and returned.

Algorithm \ref{alg:symex}  keeps track of the explored symbolic execution states in the $Active$ set. 
It associates the number of clock cycles, $numSteps$, initializes it to 0 (line 3), and updates it (line 24)
as the design is symbolically executed for one clock cycle each time (line 9). 

In Algorithm \ref{alg:symex}, our major extension to standard symbolic execution is to extract various 
types of metadata to be used in don't care transition and Trojan detection. 
One type of metadata, denoted by $Type$, is Reach, in which case it computes the set of reachable states $RS$, 
and the set of observable behaviors, $RBS$. It is important to note that in Algorithm \ref{alg:symex}, the set of 
initial states is provided as an input and so the notion of reachable is relative to the set of initial states.  
This is because we instantiate Algorithm \ref{alg:symex}, in three different ways from Algorithm \ref{alg:threestage}.
The first instantiation that performs the forward analysis uses the initial state of the design according to the standard semantics of the RTL design.
The second instantiation uses any state, denoted by $True$, as the initial state to perform the 1 clock cycle backward analysis. Finally, in the third instantiation, we use states that have been reached from an unreachable state, as the 
initial state. However, the symbolic execution state that corresponds to such a state also include any 
side effect of the Trojan trigger, e.g., the flag that enables the Trojan payload may have been set to true. 
This is because such symbolic execution states are returned as an output of the second instantiation, which 
has used the metadata type State. So, when this type of metadata type is chosen, Algorithm \ref{alg:symex} 
computes the transition relation in $Trans$ and the active set of symbolic execution states in $SymStates$. 
In most of the metadata computation the FSM state is extracted from the symbolic execution state $s$ with 
the expression $Project(s,SVar)$, where $SVar$ is the variable that implements the FSM state, e.g., {\tt pcmSq} in 
Figure \ref{fig:imafsm}.

\section{Evaluation}
\label{sec:eval}

\begin{table*}[htbp]
\caption{Summary of Don't Care Transition Detection at RTL ({\bf R}) and at Gate-Level Netlist as Generated by YOSYS ({\bf Y}) and SDC ({\bf S}).}
\label{table:dct-YOSYS}
\begin{center}
\begin{tabular}{|c|c|c|c|c|c|c|c|c|c|c|c|c|c|c|}
\hline
\multirow{2}{*}{\bf Benchmark}  & \bf {Expl.} & \multirow{2}{*}{\bf d} & 
\multicolumn{3}{|c|}{\bf $|RS|$} & 
\multicolumn{3}{|c|}{\bf $|DCT|$} & 
\multicolumn{3}{|c|}{\bf $|Dest|$} & 
\multicolumn{3}{|c|}{\bf Time (min)} \\ 
\cline{4-15}
& \bf {Method} &  & {\bf R} & {\bf Y} & {\bf S}  & {\bf R} & {\bf Y} & {\bf S} & {\bf R} & {\bf Y} & {\bf S} & {\bf R} & {\bf Y} & {\bf S} \\
\hline 

\multirow{2}{*}{\text{prep3\_binary}} 
& BFS & 6  & 8 & 8 & 8 & 0 & 0 & 0 & 0 & 0 & 0 & 58.52 & 83.47 & 15.38 \\ 
\cline{2-15}
& BFS + Pr. & 6 & 8 & 8 & 8 & 0 & 0 & 0 & 0 & 0 & 0 & 9.13 & 21.07 & 3.33\\ 
\hline

\multirow{2}{*}{\text{prep4\_binary}} 
&BFS & 8 & 16 & 16 & 16 & 0 & 0 & 0 & 0 & 0 & 0 & TO & TO & 56.08 \\ 
\cline{2-15}
& BFS + Pr. & 8 & 16 & 16 & 16 & 0 & 0 & 0 & 0 & 0 & 0 & 20.12 & 710.17 & 8.87 \\ 
\hline

\multirow{2}{*}{\text{AES128\_1}} 
&BFS & 5 & 5 & 5 & 5  & 3 & 15 & 9 & 3 & 5 & 3 & 16.25 & 68.53 & 431.42  \\ 
\cline{2-15}
&BFS + Pr. & 5 & 5 & 5 & 5  & 3 & 15 & 9 & 3 & 5 & 3 & 4.05 & 27.52 & 81.83 \\ 
\hline

\multirow{2}{*}{\text{AES128\_2}} 
&BFS & 5 & 4 & 4 & 3 & 0 & 0 & 3 & 0 & 0 & 3 & 13.72 & 14.35 & 13.01  \\
\cline{2-15}
&BFS + Pr. & 5 & 4 & 4 & 3 & 0 & 0 & 3 & 0 & 0 & 3 & 2.82 & 6.3 & 2.02 \\
\hline

\multirow{2}{*}{\text{AES128\_decrypt}} 
&BFS & 11 & 11 & 11 & 6 & 55 & 54 & 40 & 11 & 6 & 4 & 16.67 & 263.52 & 714.25 \\
\cline{2-15}
&BFS + Pr. & 11 & 11 & 11 & 6 & 55 & 54 & 40 & 11 & 6 & 4 & 4.83 & 68.41& 493.22 \\ 
\hline

\multirow{2}{*}{IMA ADPCM} 
&BFS & 7 & 6 & 6 & 6 & 2 & 12 & 12 & 1 & 6 & 6 & 6.58 & 551.55 & 616.7 \\ 
\cline{2-15}
&BFS + Pr. & 7 & 6 & 6 & 6 & 2 & 12 & 12 & 1 & 6 & 6 & 0.78 & 63.43 & 323.82  \\ 
\hline

\multirow{2}{*}{XTEA Crypto} 
&BFS & 9 & 15 & 15 & 15 & 241 & 241 & 241 & 1 & 1 & 1 & 13.08 & 621.05 & TO \\
\cline{2-15}
&BFS + Pr. & 9 & 15 & 15 & 15 & 241 & 241 & 241 & 1 & 1 & 1 & 4.82 & 390.13 & 469.37 \\
\hline

\multirow{2}{*}{APB2SPI} 
&BFS & 4 & 3 & 3 & 3 & 3 & 3 & 3 & 3 & 3 & 3 & 12.88 & 44.47 & 81.35 \\
\cline{2-15}
&BFS + Pr. & 4 & 3 & 3 & 3 & 3 & 3 & 3 & 3 & 3 & 3 & 1.53 & 11.08 & 12.65 \\ 
\hline

\multirow{2}{*}{\text{UART\_1}} 
&BFS & 5 & 7 & 1 & 2 & 7 & 7 & 12 & 7 & 1 & 2 & 64.5 & 73.72 & 84.48  \\ 
\cline{2-15}
&BFS + Pr. & 5 & 7 & 1 & 2 & 7 & 7 & 12 & 7 & 1 & 2 & 6.15 & 10.40 & 17.75 \\ 
\hline

\multirow{2}{*}{\text{UART\_2}} 
&BFS & 4 & 3 & 2 & 2 & 4 & 4 & 4 & 2 & 2 & 2 & 11.58 & 248.45 & 27.97 \\ 
\cline{2-15}
&BFS + Pr. & 4 & 3 & 2 & 2 & 4 & 4 & 4 & 2 & 2 & 2 & 1.18 & 80.35 & 12.53 \\ 
\hline

\multirow{2}{*}{\text{RS232}} 
&BFS & 3 & 5 & 2 & 4 & 15 & 12 & 12 & 5 & 2 & 3 & 2.63 & 53.62 & 5.46 \\ 
\cline{2-15}
&BFS + Pr. & 3 & 5 & 2 & 4 & 15 & 12 & 12 & 5 & 2 & 3 & 0.36 & 3.25 & 2.81 \\ 
\hline
\end{tabular}
\label{table:all}
\end{center}
\end{table*}

We have evaluated the effectiveness of our approach in terms of don't care transition detection (Section \ref{sec:contcareeval}) at RTL
(Section \ref{sec:dctatrtl}) and at gate-level (Section \ref{sec:gate-DCT}), and in terms of Trojan detection (Section \ref{sec:trojaneval}).

To evaluate our approach, we used benchmarks from \emph{OpenCores} \cite{OpenCore} and \emph{Trust-Hub} \cite{TrustHub} (except \emph{prep3} and \emph{prep4}, which we got from \cite{Golson}). These benchmarks are all described at RTL. For each benchmark, the user needs to provide the number of clock cycles for the forward analysis stage
and the name of the variable that represents the state in the FSM. 
Based on these two values a test driver is developed to execute the {\tt eval} function of the top Verilog module with symbolic inputs. 

In only two of the benchmarks ({\tt AES128} and {\tt UART}), there are more than one FSMs, e.g. {\tt AES128\_1} and {\tt AES128\_2} are two different FSMs in AES128. It is obvious that when there are multiple FSMs inside a design, which is quite common in modern circuits, the interactions between them are absolutely important. 
However, we analyze each FSM of AES128 and UART one at a time to reduce analysis complexity based on the observation that, firstly, since our analysis is performed within the context of the top module and the whole design along with making all inputs to the top module symbolic, interaction between multiple FSMs can be precisely captured while focusing on the states reached for the specific FSM. Secondly, the property we are interested in is whether all reachable states of an FSM can be explored completely and not whether certain combinations of FSM states are reachable. Lastly,
our ground truth analysis using the STGs generated 
by Intel Quartus shows that we were able to precisely detect the reachable states of a particular FSM of a design with multiple FSMs by analyzing a single FSM at a time. 

For discussions at RTL, we generated the State Transition Graph (STG) of the FSM in the RTL design under analysis using commercial software \emph{Intel Quartus} to perform ground truth analysis. 
While our symbolic execution based approach could generate the transitions in the analyzed FSMs with 100\% coverage, it could also detect don't care transitions and the Trojans hidden at such transitions efficiently and accurately.

\subsection{Detecting Don't Care Transitions}
\label{sec:contcareeval}

Although our overall approach, as shown in Figure \ref{fig:workflow}, performs don't care transition 
detection on the gate-level netlist, this 
detection can still be performed at RTL 
to help developers harden their designs.
Section \ref{sec:dctatrtl} and \ref{sec:gate-DCT} 
presents our results on detecting don't care transitions at RTL and gate-level, respectively.

\subsubsection{RTL Don't Care Transition Detection}
\label{sec:dctatrtl}

The {\bf Expl. Method} column in Table \ref{table:all} includes BFS-based full exploration and BFS pruning approaches (denoted by BFS + Pr.). The {\bf d} column represents the number of clock cycles used in the test driver to explore  the state space in forward symbolic execution. 
The next three columns, which represent the size of the set of reachable states, $|RS|$, the number of don't care transitions, $|DCT|$, and the number of unique destinations, $|Dest|$, respectively, are obtained from our method directly without referring to the STG generated by \emph{Intel Quartus}. 
The seventh column shows the time to find all DCTs in minutes and the timeout was set as three hours. 

All the don't care transitions that are detected by full exploration (BFS and BFS with Pruning, denoted by BFS +Pr.) at RTL are true positives. 
Although not reported in Table \ref{table:all}, 
we also tested the partial exploration method, which resulted in false positives in most cases (7 out of 11). This was expected based on our discussions in Section \ref{sec:twostage}.
A DCT can be a false positive either due to mislabelling of the source state as unreachable or mislabelling of the destination state as reachable, or both. 
In the partial exploration approach, the false positives are due to mislabelling of the source state as unreachable. 
This is because during the partial exploration approach, the states of the FSM may not be fully explored on some 
symbolic execution paths (or for some input sequences). Therefore, for the rest of the evaluation we use only BFS and BFS with Pruning.

\pgfplotsset{width=7.5cm,compat=1.8}
\begin{figure*}
\begin{center}
\caption{Comparison of BFS and BFS with Pruning in Don't Care Transition Detection at RTL.}
\label{fig:dctbfsvspruning}
\begin{tikzpicture}
\begin{semilogyaxis}[
    log origin=infty,
    ybar,
    bar width = 4pt,
    enlargelimits=0.15,
    legend style={at={(0.5,-0.3)},
      anchor=north,legend columns=-1},
    ylabel={DCT Time (min)},
    symbolic x coords={prep3, prep4, AES1, AES2, AESDec,IMA,XTEA,APB,UART1,UART2, RS232},
    xtick=data,
    x tick label style={rotate=45,anchor=east},
      bar width=8pt,
  height=6cm,width=15cm
]
\addplot+[ybar] coordinates {(prep3,58.52) (prep4,180) (AES1,16.25) (AES2,13.75e0) (AESDec,16.67) (IMA,9.80) (XTEA,13.08) (APB,12.88) (UART1,64.50) (UART2,11.58) (RS232,18.82)};
\addplot+[ybar] coordinates {(prep3,9.13) (prep4,20.12) (AES1,4.05) (AES2,2.82) (AESDec,4.83) (IMA,2.15) (XTEA,4.82) (APB,1.53) (UART1,6.15) (UART2,1.18) (RS232,2.8)};
\legend{\strut BFS, \strut BFS w Pruning}
\end{semilogyaxis}
\end{tikzpicture}
\end{center}
\end{figure*}
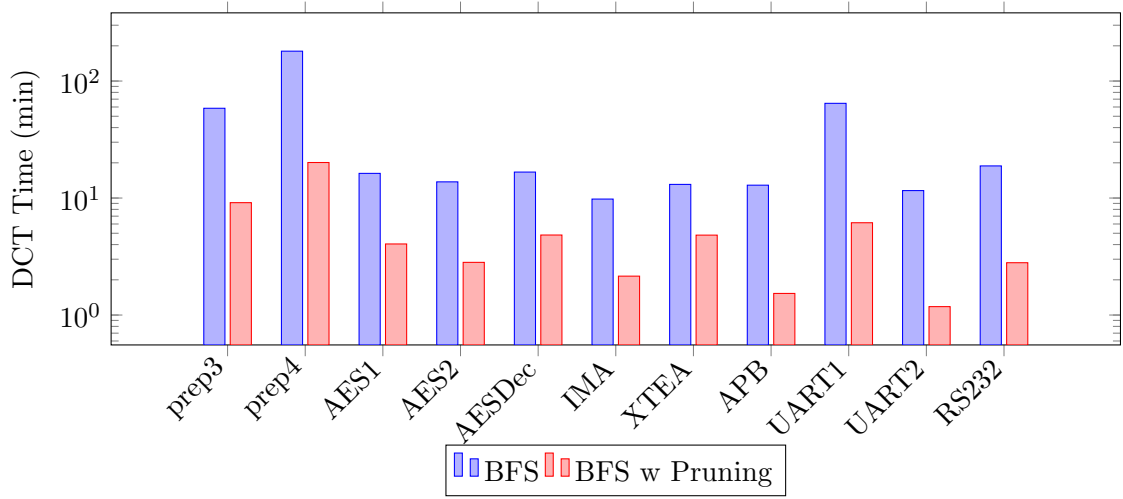

The results in Table \ref{table:all} also show the scalability and the efficiency of our approach.
As more clearly shown in Figure \ref{fig:dctbfsvspruning}, pruning the state space in the forward stage based on previously seen FSM states provides considerable speedup in don't care transition detection time. 
The average speedup is 6.86 times while maximum speedup is 10.5 times (for {\tt UART1}).
Also, for the {\tt prep4\_binary} benchmark, BFS without pruning could not report the don't care transition within 
the timeout of three hours while BFS with pruning reported those in 20.12 minutes.

\begin{figure}[htbp]
\begin{center}
\small
\includegraphics[scale=0.5]{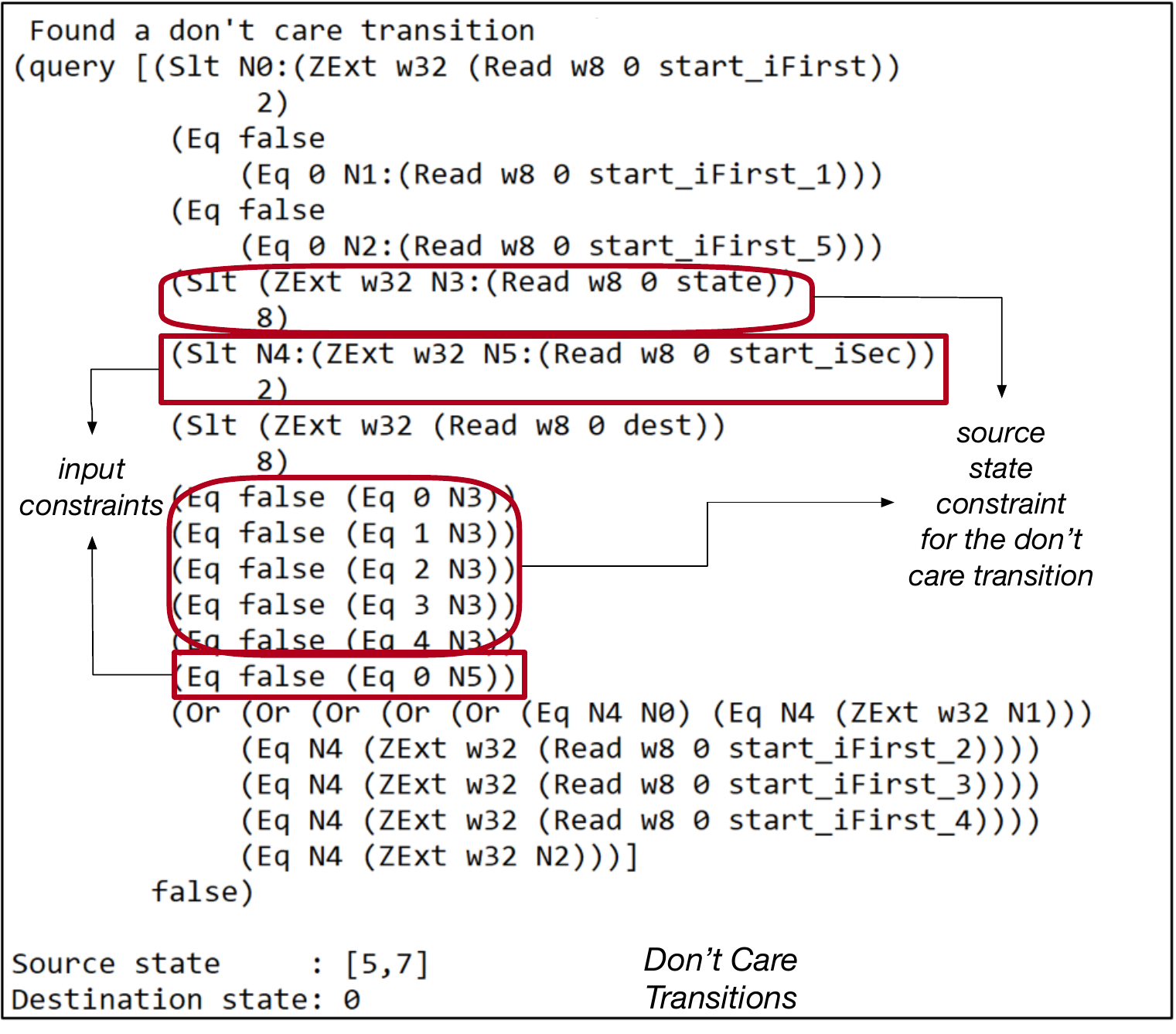}
\caption{KLEE generated symbolic constraints that describe the inputs enabling the don't care transitions in AES128\_1.}
\label{fig:testcase}
\end{center}
\end{figure}

An additional benefit of our approach is the ability to report different cases of the don't care transitions along with the 
input conditions.
For example,  Figure \ref{fig:testcase} presents a constraint that is generated by the symbolic execution 
engine, which indicates that 
 under the input condition {\tt start\_iSec} $<$ 2 and {\tt start\_iSec} $\not =$ 0 (see the constraints inside 
 the rectangles in Figure \ref{fig:testcase}), {\tt state} could transit from {\tt state} $<$ 8 and {\tt state} $\neq$ 0,1,2,3,4 (see the constraints inside rounded rectangles in Figure \ref{fig:testcase}), which means
the don't care transition can be from source {\tt state  = 5, 6, or 7}, to destination state 0. 
So, besides reporting don't care transitions, our method can also report the don't care states (unspecified states), 
 state 5, 6 and 7 in this case.

In summary, the analysis time is determined by three aspects. First, the time to detect the don't care transitions become longer as the design gets more complicated. Second, as the number of clock cycles (or {\bf d}) increases, symbolic execution may generate exponentially more paths. 
Finally, the pruning technique can speedup the detection of don't care transitions by eliminating redundant paths with respect to the FSM states.

\subsubsection{Gate-Level Don't Care Transition Detection}
\label{sec:gate-DCT}

Don't care transitions in IPs can be introduced by careless designers, synthesis tools, or both. In this section, two synthesis tools, an open-source software, Yosys Open Synthesis Suite (YOSYS) \cite{YOSYS}, and a  commercial software, Synopsys Design Compiler (SDC) \cite{DC}, are used to translate RTL designs into the corresponding gate-level netlists.
Then we perform our Trojan detection approach on those netlists, and compare these two tools with each other as well as their results with those we obtained from RTL.

YOSYS  is an open-source framework for Verilog RTL synthesis. It can take behavioral design description as input and generate gate-level description, either logical gate or physical gate, of the design as output \cite{YOSYS}. Thus, in our workflow, gate-level Verilog is first generated by providing a simple script and the remaining steps are the same as that of RTL Verilog (i.e. generate C++ model of gate-level Verilog along with user-provided test harness, compile C++ code, extract LLVM bitcode, and perform symbolic execution).

SDC can automatically synthesize an optimized gate-level circuit based on the design description and design constraints. It can also accept multiple input formats and generate structure Verilog along with multiple performance reports. Workflow in SDC DCT approach is actually same with that of YOSYS DCT approach, except that synthesized gate-level netlist is obtained from \emph{Design\_Ultra} with \emph{saed90nm} target library and symbol library.

\begin{figure}[htbp]
\begin{center}
\includegraphics[scale=0.5]{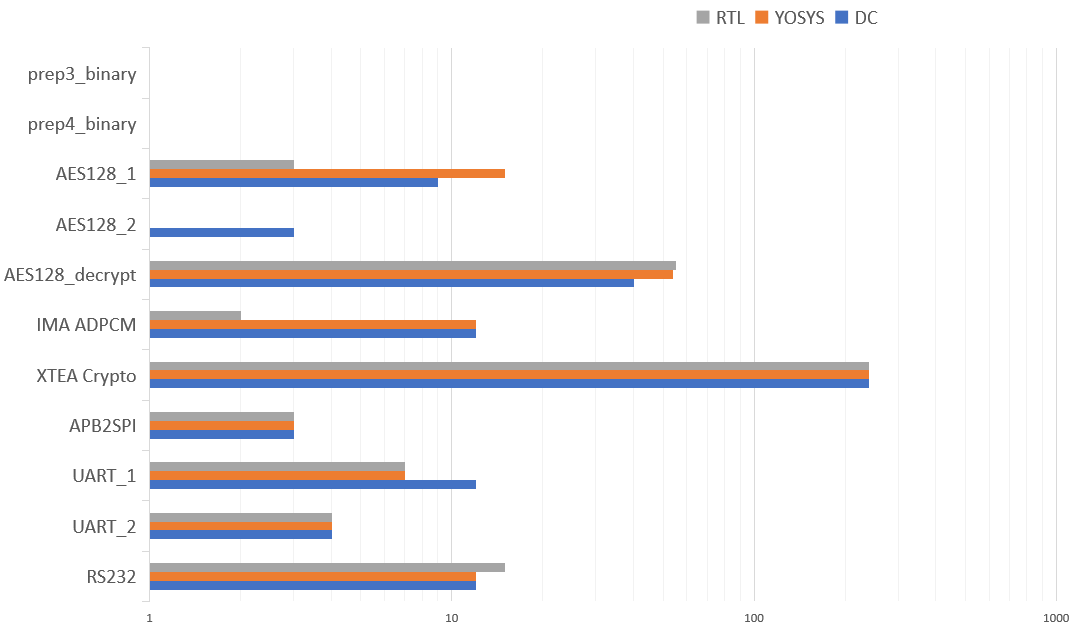}
\caption{\# of DCTs of RTL HDL, YOSYS- and Synopsys Design Compiler-generated gate-level netlist.}
\label{fig:dctCompare}
\end{center}
\end{figure}

Table \ref{table:dct-YOSYS} lists the summary of DCT detection at RTL and at gate-level netlist as generated by YOSYS and SDC. 
As Figure \ref{fig:dctCompare} shows for most of 
the benchmarks the number of DCTs at RTL is same 
as those at gate-level netlist.
However, there are some cases for which 
the synthesis tool introduces additional 
DCTs.
For example, in two (\emph{AES128\_1}, \emph{IMA APDCM}) out of 11 benchmarks, 
the number of DCTs in the YOSYS-generated gate-level netlist is larger than that of RTL. 
Similarly, in four (\emph{AES128\_1}, \emph{AES128\_2}, \emph{IMA APDCM}, \emph{UART\_1}) out of 11 benchmarks, the number of DCTs in the SDC generated netlist is larger than that of RTL. 
On the other hand, in some cases, the synthesis tools may also reduce the number of DCTs.
For example, in two (\emph{AES128\_decrypt} and 
\emph{RS232}) out of 11 benchmarks, 
the number of DCTs in YOSYS-generated and SDC-generated gate-level netlists are smaller 
than that of RTL. However, for \emph{AES128\_decrypt}, SDC has a bigger DCT reduction (15) compared to what is achieved by YOSYS (1). 
Also, such reductions may be due to the optimization of the FSM as in the 
case of \emph{RS232}, which is indicated by the reduction in the set of reachable states in the gate-level netlist as shown in Table \ref{table:all}.

\begin{figure}[htbp]
\begin{center}
\includegraphics[scale=0.45]{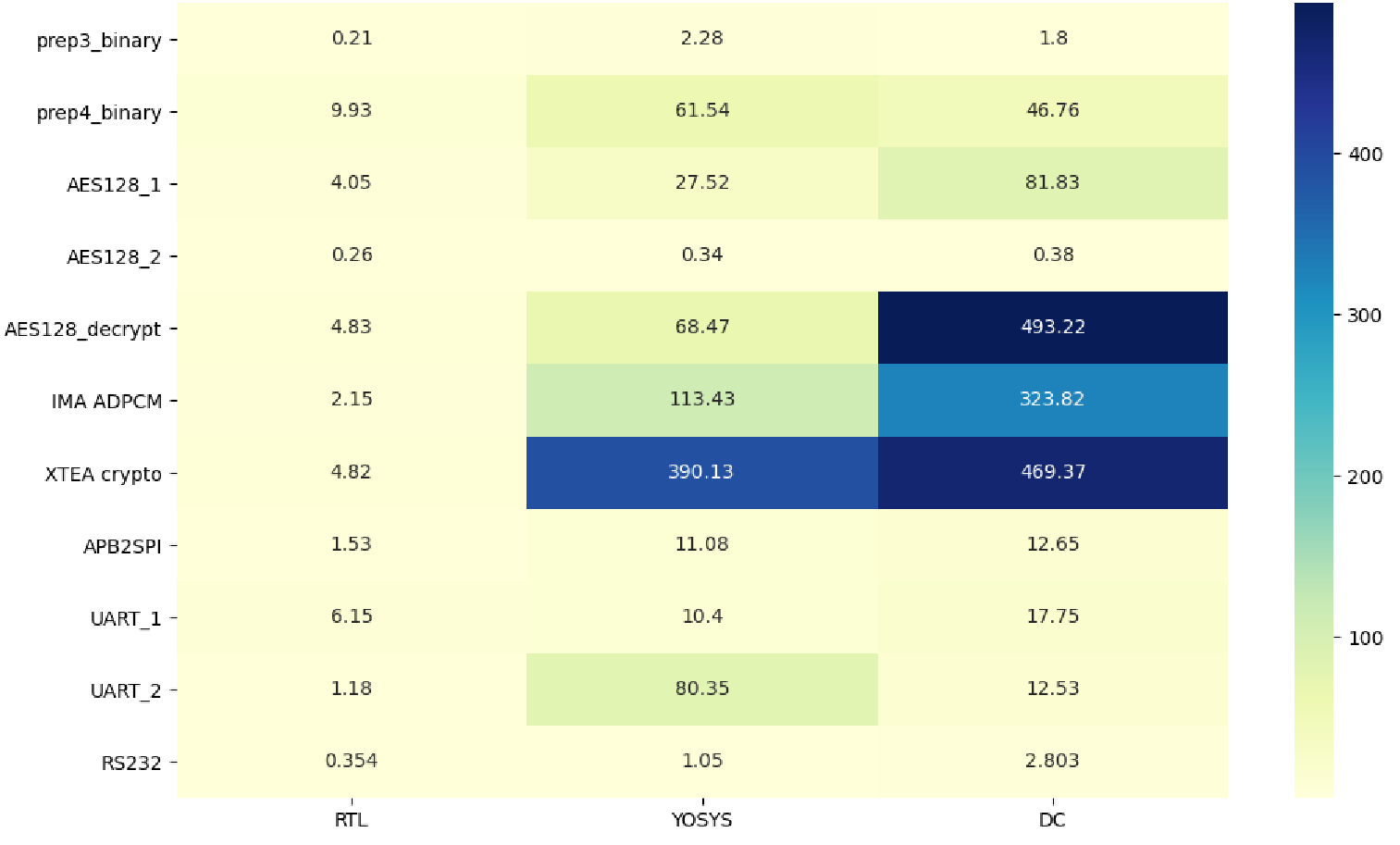}
\caption{The heat map for the total run time with pruning technique for RTL HDL, YOSYS- and Design Compiler-generated gate-level netlist. All times are reported in minutes, and timeout is 600 minutes.}
\label{fig:heatmap_runtime}
\end{center}
\end{figure}

\begin{figure}[htbp]
\begin{center}
\includegraphics[scale=0.75]{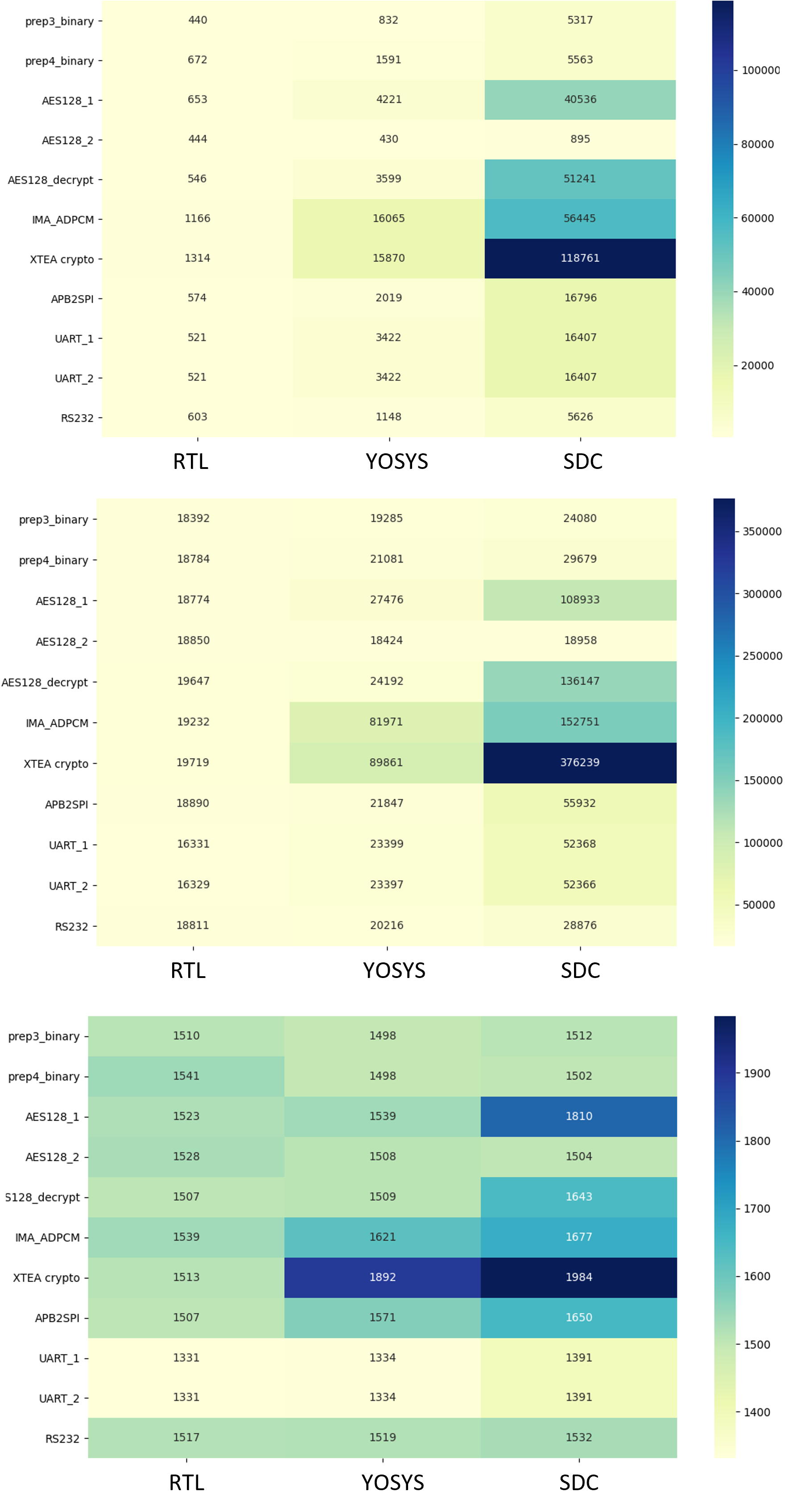}
\caption{SW LOC, ICounts, and BCounts (from top to bottom respectively) of RTL netlist, YOSYS-generated netlist, and Design Compiler-generated netlist.}
\label{fig:heatmap_3in1}
\end{center}
\end{figure}

To demonstrate effectiveness of performing DCT detection at RTL, run times of each benchmarks 
regarding RTL, YOSYS- and SDC-generated gate-level netlist are shown by a heatmap  in Figure \ref{fig:heatmap_runtime}. For YOSYS-generated netlist, run times of all gate-level netlist benchmarks increase at most by 81 times (\emph{IMA ADPCM}) compared with that of RTL. Similarly, for all gate-level netlist synthesized by SDC, run times increase at most by 415 times (\emph{IMA ADPCM}) compared with that of RTL. 

Additionally, to explain huge runtime increase at gate-level netlist, we report the complexity of each benchmark in software domains, in terms of Software Lines of Code (SW LOC), LLVM instruction count (ICounts), and LLVM branch instruction count (BCounts). As more clearly shown in Fig.\ref{fig:heatmap_3in1},
heatmaps of each complexity criteria are provided. Although it's infeasible to determine complexity by a certain criteria, these criteria can be used as indicators to reveal potential complexity. For example, for \emph{IMA\_ADPCM}, \emph{XTEA Crypto} in both YOSYS- and SDC-generated netlist, \emph{AES128\_decrypt} in SDC-generated netlist, individually each of the three criteria (SW LOC, ICounts, and BCounts) is much larger than that of the other benchmarks, which corresponds to a much larger increase in runtime as shown in Fig.\ref{fig:heatmap_runtime}.

Although it's desirable to perform DCT detection approach at the RTL level for performance reasons, {\em for soundness DCTs must be detected 
at the gate-level to account for the DCTs that may be introduced by the synthesis tools}.

\subsection{Detecting Trojans Hidden at Don't Care Transitions}
\label{sec:trojaneval}

We evaluate our Trojan detection approach on
two case studies: the {\tt RS232-T700} benchmark from 
{\bf TrustHub} \cite{TrustHub} and the {\tt IMA ADPCM} audio encoding algorithm \cite{IMA_ADPCM}.
Since none of the Trojans in the TrustHub \cite{TrustHub} benchmarks use a don't care transition as a trigger, we decided to use existing Trojan payloads from TrustHub and identify RTL designs with FSMs that have don't care transitions so that we could generate new Trojan benchmarks suitable for our approach.
So, it turned out that the {RS232-T700} benchmark  from TrustHub\cite{TrustHub}, specifically, the transmitter part in {RS232-T700}, satisfied all our requirements perfectly. Besides, we choose IMA ADPCM benchmark from Opencores \cite{OpenCore} for the reason that, both synthesis tools (YOSYS and SDC) could introduce same additional DCTs with the same reachable states, which supports the discussion in Section \ref{sec:gatetrojaneval}.
We evaluate Trojan detection for the two case studies in Sections \ref{sec:ima} and \ref{sec:gatetrojaneval}.

\subsubsection{Trojan Detection for IMA ADPCM}
\label{sec:ima}

In this section, we apply the DCT and Trojan detection method to the encoder part of the IMA ADPCM benchmark. The IMA ADPCM audio compression algorithm belongs to the Adaptive Differential Pulse Code Modulation type algorithms. The algorithm is based on a simple adaptive quantization of the difference between the input and predictor. Each 16-bit input sample is converted to a 4-bit coded
information which yields a compression ratio of ¼\cite{IMA_ADPCM}. We will not go through detailed implementation of the design, instead, we focus on the FSM part, which is shown in Figures  \ref{fig:imafsm} and \ref{fig:fsmex}. 

As summarized in Table \ref{table:dct-YOSYS}, for \emph{IMA\_ADPCM} benchmark, reachable states at RTL, YOSYS-generated netlist, and  SDC-generated netlist are the same (Column {\bf $|RS|$}). However, \# DCTs increases after synthesis (Column {\bf $|DCT|$}), which means synthesis tools could produce additional DCTs. Trojans that utilize DCTs that are introduced after the synthesis would not be detected if Trojan analysis only checks for DCTs that are detectable at RTL. Therefore, we perform DCT detection at gate-level. However, for this benchmark, since the synthesis preserves the 
FSM, we can perform Trojan detection, specifically, the divergent output detection stage in Figure \ref{fig:workflow}, at either RTL or gate-level. 

\begin{table}[htbp]
\caption{Results on Trojans Triggered by DCTs that Exist at the Synthesised Gate-level Netlist for  IMA ADPCM.}
\label{tab:addtiondctTroj}
\begin{center}
\begin{tabular}{|c|c|c|c|c|c|c|}
\hline
\multirow{2}{*}{\bf Trojan}  &
\multirow{2}{*}{\bf DCTs}  &
\multicolumn{2}{|c|}{\bf DCT Time (min)} & 
\multicolumn{3}{|c|}{\bf Trojan Time (min)}  \\ 
\cline{3-7}
 &  & YOSYS & SDC  & RTL & YOSYS & SDC  \\
\hline 

ima-1 & {T(6$\rightarrow$0)} & 66.61 & 321.79 & 1.377 & 155.52 & 476.92 \\
\hline 

ima-2 & \textcolor{blue}{T(6$\rightarrow$1)} & 67.83 & 329.43 & {0.52} & 159.47 & 480.33 \\
\hline 

ima-3 & \textcolor{blue}{T(6$\rightarrow$2)} & 66.13 & 318.74 & {0.538} & 160.35 & 485.67 \\
\hline 

ima-4 & \textcolor{blue}{T(6$\rightarrow$3)} & 66.56 & 320.63 & {0.928} & 152.93 & 474.71 \\
\hline

ima-5 & \textcolor{blue}{T(6$\rightarrow$4)} & 67.31 & 335.41 & {0.851} & 158.62 & 480.43 \\
\hline

ima-6 & \textcolor{blue}{T(6$\rightarrow$5)} & 64.83 & 328.18 & {0.819} & 156.67 & 477.36 \\
\hline

ima-7 & {T(7$\rightarrow$0)} & 69.15 & 338.23 & 1.338 & 160.56 & 482.94 \\
\hline

ima-8 & \textcolor{blue}{T(7$\rightarrow$1)} & 66.27 & 332.38 & {0.587} & 153.94 & 470.83 \\
\hline

ima-9 & \textcolor{blue}{T(7$\rightarrow$2)} & 67.79 & 337.61 & 
{0.598} & 156.83 & 475.31 \\
\hline

ima-10 & \textcolor{blue}{T(7$\rightarrow$3)} & 68.71 & 329.17 & 
{0.951} & 158.20 & 479.64 \\
\hline

ima-11 & \textcolor{blue}{T(7$\rightarrow$4)} & 67.34 & 339.56 & 
{0.892} & 156.89 & 484.13 \\
\hline

ima-12 & \textcolor{blue}{T(7$\rightarrow$5)} & 67.58 & 332.49 & 
{0.871} & 157.41 & 487.67 \\
\hline
\end{tabular}
\end{center}
\end{table}

Thus, twelve different DCT-triggered Trojan versions of \emph{IMA APDCM} based on each DCT obtained from analysis at the gate-level netlist are designed as shown in Table \ref{tab:addtiondctTroj}.
It is interesting to note that, in this table, only \textbf{ima-1} and \textbf{ima-7}  represent Trojans triggered by DCTs that only exist at RTL and the remaining ten versions (in blue) indicate Trojans triggered by additional DCTs introduced by the synthesis tool, all activated Trojans will make the output \emph{outValid} stuck at '1'.

As the timing results in Table \ref{tab:addtiondctTroj} indicate, 
Trojan detection at RTL is much faster than the one at gate-level.
Specifically, Trojan detection time at 
gate-level netlist is 
 on average 116.5 times bigger for YOSYS-generated 
 netlist and 353.6 times bigger for SDC-generated 
 netlist than that of Trojan detection at RTL. 
 Also, Trojan detection at gate-level takes much longer for SDC-generated netlists compared to 
 those for YOSYS-generated netlist, which is 
 shown more clearly in Figure \ref{fig:addDCTTroj}. 
 This can be explained in terms of the 
 code complexity difference; as shown in Figure 
 \ref{fig:heatmap_3in1} SDC-generated netlists are 
 in general more complex than YOSYS-generated netlists.
 
 As shown in Figure \ref{fig:yosysvsrtltrojtime} and Figure \ref{fig:dcvsrtltrojtime}, using the hybrid approach of detecting DCTs at gate-level and detecting the Trojans at RTL, we can achieve an average of 3.3X speed-up for YOSYS-generated netlists and an average of 2.46X speed-up for SDC generated netlist.

\pgfplotsset{compat=1.16}
\begin{figure*}[th!]
\begin{center}
\caption{Comparison of DCT and Trojan Detection Times on the Synthesized Gate-level Netlists (YOSYS vs SDC generated) for IMA ADPCM. The Trojans 
are Triggered by DCTs that Exist at the Synthesised Gate-level Netlist.}
\label{fig:addDCTTroj}
\begin{tikzpicture}[
  every axis/.style={ 
    ybar stacked,
    ybar=10pt,
    ymin=-1,ymax=1000,
    x tick label style={rotate=45,anchor=east},
    ylabel={Total Time (min)},
    symbolic x coords={
      ima-1, ima-2, ima-3,ima-4,ima-5, ima-6, ima-7, ima-8, ima-9, ima-10, ima-11, ima-12
    },
  bar width=8pt,
  xtick=data,
  height=6cm,width=15cm
  },
]

\begin{semilogyaxis}[bar shift=-8pt, hide axis, transpose legend,legend columns=2, legend style={at={(0.3,-0.25)},anchor=north}]
\addplot coordinates
{(ima-1,66.61) (ima-2,67.83) (ima-3,66.13) (ima-4,66.56) (ima-5,67.31) (ima-6,64.83) (ima-7,69.15) (ima-8,66.27) (ima-9,67.79) (ima-10,68.71) (ima-11,67.34) (ima-12,67.58)};
\addlegendentry{YOSYS-DCT}
\addplot coordinates
{(ima-1,155.52) (ima-2,157.73) (ima-3,153.6) (ima-4,152.93) (ima-5,158.62) (ima-6,156.67) (ima-7,160.56) (ima-8,153.94) (ima-9,156.83) (ima-10,158.2) (ima-11,156.89) (ima-12,157.41)};
\addlegendentry{YOSYS-Troj}
\end{semilogyaxis}

\begin{semilogyaxis}[bar shift=1.5pt, transpose legend,legend columns=2, legend style={at={(0.7,-0.25)},anchor=north}]
\addplot+[fill=blue!50!gray] coordinates
{(ima-1,321.79) (ima-2,329.43) (ima-3,318.74) (ima-4,320.63)  (ima-5,335.41) (ima-6,328.18) (ima-7,338.23) (ima-8,332.38) (ima-9,329.17) (ima-10,337.61) (ima-11,339.56) (ima-12,332.49)};
\addlegendentry{SDC-DCT}
\addplot+[fill=red!50!gray] coordinates
{(ima-1,476.92) (ima-2,480.33) (ima-3,485.67) (ima-4,474.41) (ima-5,480.43) (ima-6,477.36) (ima-7,482.94) (ima-8,470.83) (ima-9,475.31) (ima-10,479.64) (ima-11,484.13) (ima-12,487.67)};
\addlegendentry{SDC-Troj}
\end{semilogyaxis}
\end{tikzpicture}
\end{center}
\end{figure*}

\pgfplotsset{compat=1.8}
\begin{figure*}[th!]
\begin{center}
\caption{Comparison of YOSYS-Generated Gate-Level Netlist DCT Detection time + RTL Trojan Detection Time vs YOSYS-Generated Gate-Level Netlist DCT + Trojan Detection Time  on IMA\_ADPCM for Trojans that are Triggered by DCTs that Exist at the Gate-Level Netlist.}
\label{fig:yosysvsrtltrojtime}
\begin{tikzpicture}[
  every axis/.style={ 
    ybar stacked,
    ybar=10pt,
    ymin=0,ymax=800,
    x tick label style={rotate=45,anchor=east},
    ylabel={Total Time (min)},
    symbolic x coords={
      ima-1, ima-2, ima-3,ima-4,ima-5, ima-6, ima-7, ima-8, ima-9, ima-10, ima-11, ima-12
    },
  bar width=8pt,
  xtick=data,
  height=6cm,width=15cm
  },
]

\begin{semilogyaxis}[bar shift=-8pt, hide axis, transpose legend,legend columns=2, legend style={at={(0.3,-0.25)},anchor=north}]
\addplot coordinates
{(ima-1,66.61) (ima-2,67.83) (ima-3,66.13) (ima-4,66.56) (ima-5,67.31) (ima-6,64.83) (ima-7,69.15) (ima-8,66.27) (ima-9,67.79) (ima-10,68.71) (ima-11,67.34) (ima-12,67.58)};
\addlegendentry{YOSYS-DCT}
\addplot coordinates
{(ima-1,1.377) (ima-2,0.52) (ima-3,0.538) (ima-4,0.928) (ima-5,0.851) (ima-6,0.819) (ima-7,1.338) (ima-8,0.587) (ima-9,0.598) (ima-10,0.951) (ima-11,0.892) (ima-12,0.871)};
\addlegendentry{RTL-Troj}
\end{semilogyaxis}

\begin{semilogyaxis}[bar shift=1.5pt, transpose legend,legend columns=2, legend style={at={(0.7,-0.25)},anchor=north}]
\addplot+[fill=blue!50!gray] coordinates
{(ima-1,66.61) (ima-2,67.83) (ima-3,66.13) (ima-4,66.56) (ima-5,67.31) (ima-6,64.83) (ima-7,69.15) (ima-8,66.27) (ima-9,67.79) (ima-10,68.71) (ima-11,67.34) (ima-12,67.58)};
\addlegendentry{YOSYS-DCT}
\addplot+[fill=red!50!gray] coordinates
{(ima-1,155.52) (ima-2,157.73) (ima-3,153.6) (ima-4,152.93) (ima-5,158.62) (ima-6,156.67) (ima-7,160.56) (ima-8,153.94) (ima-9,156.83) (ima-10,158.2) (ima-11,156.89) (ima-12,157.41)};
\addlegendentry{YOSYS-Troj}
\end{semilogyaxis}
\end{tikzpicture}
\end{center}
\end{figure*}

\pgfplotsset{compat=1.8}
\begin{figure*}[th!]
\begin{center}
\caption{Comparison of SDC-Generated Gate-Level Netlist DCT Detection time + RTL Trojan Detection Time vs SDC-Generated Gate-Level Netlist DCT + Trojan Detection Time  on IMA\_ADPCM for Trojans that are Triggered by DCTs that Exist at the Gate-Level Netlist.}
\label{fig:dcvsrtltrojtime}
\begin{tikzpicture}[
  every axis/.style={ 
    ybar stacked,
    ybar=10pt,
    ymin=-1,ymax=1000,
    x tick label style={rotate=45,anchor=east},
    ylabel={Total Time (min)},
    symbolic x coords={
      ima-1, ima-2, ima-3,ima-4,ima-5, ima-6, ima-7, ima-8, ima-9, ima-10, ima-11, ima-12
    },
  bar width=8pt,
  xtick=data,
  height=6cm,width=15cm
  },
]

\begin{semilogyaxis}[bar shift=-8pt, hide axis, transpose legend,legend columns=2, legend style={at={(0.3,-0.25)},anchor=north}]
\addplot coordinates
{(ima-1,321.79) (ima-2,329.43) (ima-3,318.74) (ima-4,320.63)  (ima-5,335.41) (ima-6,328.18) (ima-7,338.23) (ima-8,332.38) (ima-9,329.17) (ima-10,337.61) (ima-11,339.56) (ima-12,332.49)};
\addlegendentry{SDC-DCT}
\addplot coordinates
{(ima-1,1.377) (ima-2,0.52) (ima-3,0.538) (ima-4,0.928) (ima-5,0.851) (ima-6,0.819) (ima-7,1.338) (ima-8,0.587) (ima-9,0.598) (ima-10,0.951) (ima-11,0.892) (ima-12,0.871)};
\addlegendentry{RTL-Troj}
\end{semilogyaxis}

\begin{semilogyaxis}[bar shift=1.5pt, transpose legend,legend columns=2, legend style={at={(0.7,-0.25)},anchor=north}]
\addplot+[fill=blue!50!gray] coordinates
{(ima-1,321.79) (ima-2,329.43) (ima-3,318.74) (ima-4,320.63)  (ima-5,335.41) (ima-6,328.18) (ima-7,338.23) (ima-8,332.38) (ima-9,329.17) (ima-10,337.61) (ima-11,339.56) (ima-12,332.49)};
\addlegendentry{SDC-DCT}
\addplot+[fill=red!50!gray] coordinates
{(ima-1,476.92) (ima-2,480.33) (ima-3,485.67) (ima-4,474.41) (ima-5,480.43) (ima-6,477.36) (ima-7,482.94) (ima-8,470.83) (ima-9,475.31) (ima-10,479.64) (ima-11,484.13) (ima-12,487.67)};
\addlegendentry{SDC-Troj}
\end{semilogyaxis}
\end{tikzpicture}
\end{center}
\end{figure*}
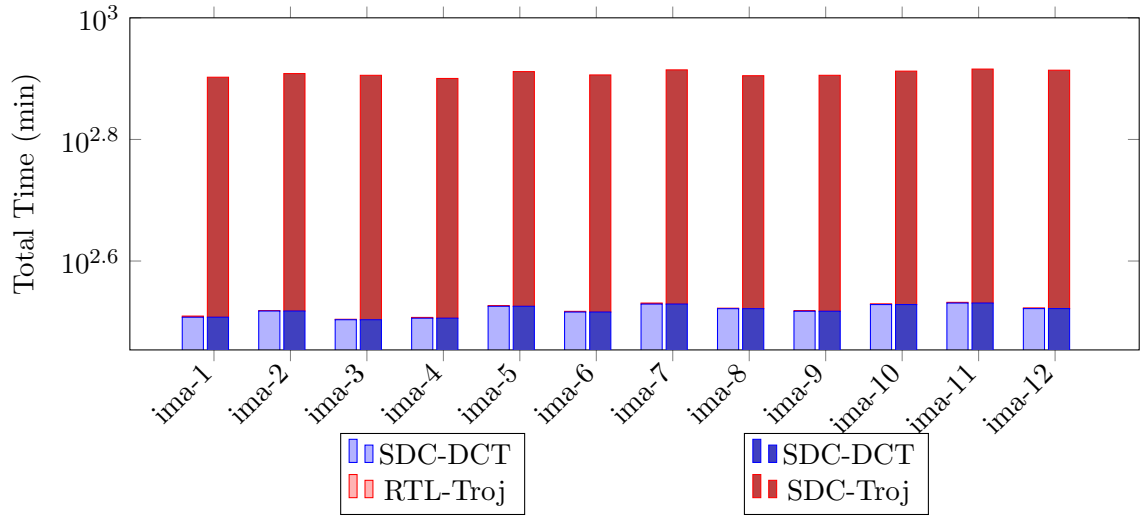

\begin{table}[th!]
\caption{Results on Trojans Triggered by DCTs that Exist at the Synthesized Gate-level Netlist for
RS232-T700.}
\label{table:gaters232trojananalysis}
\begin{center}
\begin{tabular}{|c|c|c|c|c|c|}
\hline
\multicolumn{6}{c}{\bf YOSYS-generated netlist}\\
\hline
\hline
\multirow{2}{*}{{\bf Trojan}} & \multirow{2}{*}{{\bf Trigger}}     & \multirow{2}{*}{{\bf d}} & {\bf DCT time} & {\bf Trojan time}  &{\bf Total time} \\
&   & & {\bf (min)} & {\bf (min)} & {\bf (min)} \\
\hline 
{RS232-T1} & {T(1$\rightarrow$0)} & {5} & {8.58} & {2.18} & {10.76}\\
\hline
{RS232-T2} & {T(1$\rightarrow$2)} & {5} & {8.32} & {2.01} & {10.33}\\
\hline
{RS232-T3} & {T(3$\rightarrow$0)} & {5} & {8.93} & {2.31} & {11.24} \\
\hline
{RS232-T4} & {T(3$\rightarrow$2)} & {5} & {9.13} & {2.44} & {11.57}\\
\hline
{RS232-T5} & {T(4$\rightarrow$0)} & {5} & {8.53} & {2.39} & {10.92}\\
\hline
{RS232-T6} & {T(4$\rightarrow$2)} & {5} & {8.88} & {2.62} & {11.50}\\
\hline
{RS232-T7} & {T(5$\rightarrow$0)} & {5} & {9.15} & {2.61} & {11.76}\\
\hline
{RS232-T8} & {T(5$\rightarrow$2)} & {5} & {9.02} & {2.45} & {11.56}\\
\hline
{RS232-T9} & {T(6$\rightarrow$0)} & {5} & {8.38} & {2.21} & {10.59}\\
\hline
{RS232-T10} & {T(6$\rightarrow$2)} & {5} & {8.78} & {2.37} & {11.15}\\
\hline
{RS232-T11} & {T(7$\rightarrow$0)} & {5} & {8.34} & {2.08} & {10.42}\\
\hline
{RS232-T12} & {T(7$\rightarrow$2)} & {5} & {8.95} & {2.28} & {11.23}\\
\hline
\hline
\multicolumn{6}{c}{\bf SDC-generated netlist}\\
\hline
\hline
\multirow{2}{*}{{\bf Trojan}} & \multirow{2}{*}{{\bf Trigger}}     & \multirow{2}{*}{{\bf d}} & {\bf DCT time} & {\bf Trojan time}  &{\bf Total time} \\
&   & & {\bf (min)} & {\bf (min)} & {\bf (min)} \\
\hline 
{RS232-T1} & {T(4$\rightarrow$0)} & {5} & {6.59} & {7.67} & {14.26}\\
\hline
{RS232-T2} & {T(4$\rightarrow$2)} & {5} & {6.16} & {7.13} & {13.29}\\
\hline
{RS232-T3} & {T(4$\rightarrow$3)} & {5} & {6.85} & {8.04} & {14.89}\\
\hline
{RS232-T4} & {T(5$\rightarrow$0)} & {5} & {6.44} & {7.55} & {13.99}\\
\hline
{RS232-T5} & {T(5$\rightarrow$2)} & {5} & {7.13} & {8.29} & {15.42}\\
\hline
{RS232-T6} & {T(5$\rightarrow$3)} & {5} & {6.73} & {7.91} & {14.64}\\
\hline
{RS232-T7} & {T(6$\rightarrow$0)} & {5} & {6.80} & {7.95} & {14.75}\\
\hline
{RS232-T8} & {T(6$\rightarrow$2)} & {5} & {6.96} & {8.09} & {15.05}\\
\hline
{RS232-T9} & {T(6$\rightarrow$3)} & {5} & {7.15} & {8.35} & {15.50}\\
\hline
{RS232-T10} & {T(7$\rightarrow$0)} & {5} & {6.90} & {7.98} & {14.88}\\
\hline
{RS232-T11} & {T(7$\rightarrow$2)} & {5} & {6.26} & {7.33} & {13.59}\\
\hline
{RS232-T12} & {T(7$\rightarrow$3)} & {5} & {6.94} & {8.08} & {15.02}\\
\hline
\end{tabular}
\end{center}
\end{table}

\subsubsection{Trojan Detection for RS232-T700}
\label{sec:gatetrojaneval}

We modified the RS232-T700 benchmark to generate twelve benchmarks; one for each DCT, and named each one as {\tt RS232-T700-N}, where $N$ represents the DCT number. Since the synthesis process does not 
preserve the FSM for this benchmark, as indicated by the columns under $|RS|$ in Table \ref{table:all}, 
Trojan detection can soundly be performed only at the gate-level for this benchmark. 

As the results in Table \ref{table:gaters232trojananalysis} show, 
the total analysis time (DCT + Trojan detection) 
takes at most 15.50 mins and Trojan detection at gate-level takes at most 8.35 mins. This is in contrast with the analysis times for the IMA\_ADPCM benchmark, for which gate-level Trojan detection takes at the minimum 152.93 mins. 
This can be explained with the relative code complexity of these benchmarks: as shown in Figure \ref{fig:heatmap_3in1}, YOSYS and SDC-generated 
gate-level netlists for IMA\_ADPCM are more 
complex than those for RS232-T700.

\section{Discussion}
\label{sec:discussion}

Our approach has limitations due to several assumptions we make, the type of deviant behavior that we can handle, and the type of benchmarks we used. We assume that we know the diameter of the FSM of interest and the variable (register) that represents the state as our approach expects these as inputs. 
Although we were able to easily identify these inputs 
precisely for the benchmark designs, failing to precisely specify these  
as input would yield false positives as well as false negatives in Trojan detection. 
This can happen, for instance, by setting the number of steps to a value 
that is smaller than the diameter or by not recognizing all registers that represent the states of an FSM, e.g., the combined values of multiple registers represent the state and only a strict subset of these  registers are provided as input.

The second type of limitation is due to detecting 
deviant behavior in terms of output values. If the 
Trojan causes other types of side effects such as 
leaking information through side channels, 
our approach cannot detect such Trojans. However, 
our don't care transition detection would still be effective and can guide other approaches that can reason about side channels.

The third type of limitation is due to restricting our analysis to IP-level designs. Although using our 
pruning techniques we can achieve significant 
performance improvement, our approach may face the path explosion problem and needs to 
be improved to scale it to the detection of SOC-level Trojan analysis. 

\section{Conclusions}
\label{sec:conc}

We presented a don't care transition and Trojan detection technique that does not require a golden design or a specification.
Our approach leverages symbolic execution to achieve high coverage of the
reachable and unreachable states in an FSM. 
Our three stage approach extracts transitions  of an FSM and detects don't care transitions, 
and the Trojans hidden at such transitions. 
We developed a state space pruning technique and applied our approach  
to various benchmarks from OpenCores and Trust-hub.
Our results show that our approach achieves 100\% precision while our pruning technique speeds up detection of the don't care transitions up to 10 times. 
We show that while don't care transitions must be 
detected at gate-level, Trojan detection can 
be performed at RTL provided that the synthesis preserves the FSM structure. 
In future work, we will investigate scaling our approach to SoC designs and to the detection of Trojans with additional types of payloads.

\section*{Acknowledgements}
This project has been partially funded by NSF Award 2019283. We would like to thank Yier Jin, Xiaolong Guo, and 
Orlando Arias for the valuable discussions about the synthesis process.

\bibliographystyle{plain}
\bibliography{dontcare.bib}

\end{document}